\documentclass{emulateapj}

\usepackage{graphicx}
\usepackage{float}
\usepackage{amsmath}
\usepackage{epsfig,floatflt}

\begin{document}

\title{Directional variations of the non-Gaussianity parameter $\MakeLowercase{f}_{NL}$}

\author{
\O{}ystein Rudjord\altaffilmark{1,2,3},
    Frode K. Hansen\altaffilmark{2,3},
    Xiaohong Lan\altaffilmark{4}\\
    Michele Liguori\altaffilmark{5}
    Domenico Marinucci\altaffilmark{4}
    Sabino Matarrese\altaffilmark{6}}

  \altaffiltext{1}{email: oystein.rudjord@astro.uio.no}
  \altaffiltext{2}{Institute of Theoretical Astrophysics, University
    of Oslo, P.O.\ Box 1029 Blindern, N-0315 Oslo, Norway}
  \altaffiltext{3}{Centre of Mathematics for Applications, University of
    Oslo, P.O.\ Box 1053 Blindern, N-0316 Oslo, Norway}
  \altaffiltext{4}{Dipartimento di Matematica, Universit\`a di Roma
    `Tor Vergata', Via della Ricerca Scientifica 1, I-00133 Roma,
    Italy} \altaffiltext{5}{Department of Applied Mathematics and
    Theoretical Physics, Centre for Mathematical Sciences, University
    of Cambridge, Wilberfoce Road, Cambridge, CB3 0WA, United Kingdom}
  \altaffiltext{6}{Dipartimento di Fisica, G. Galilei, Universit\`a di
    Padova and INFN, Sezione di Padova,via Marzolo 8,I-35131 Padova,
    Italy}

%\date{Received - / Accepted -}

\begin{abstract}

  We investigate local variations of the primordial
  non-Gaussianity parameter $f_{NL}$ in the WMAP data, looking for possible influence of foreground contamination in the full-sky estimate of $f_{NL}$. We first improve the needlet bispectrum estimate in ( Rudjord et al. 2009) on the full-sky to $f_{NL}=73\pm31$  using the KQ75 mask on the co-added V+W channel. We find no particular values of $f_{NL}$ estimates close to the galactic plane and conclude that foregrounds are unlikely to affect the estimate of $f_{NL}$ in the V and W bands even for the smaller KQ85 mask. In the Q band however, we find unexpectedly high values of $f_{NL}$ in local estimates close to the galactic mask, as well as significant discrepancies between Q band estimates and V/W band estimates. We therefore conclude that the Q band is too contaminated to be used for non-Gaussianity studies even with the larger KQ75 mask.
We further noted that the local $f_{NL}$ estimates on the V+W channel are positive on all equatorial bands from the north to the south pole. The probability for this to happen in a universe with $f_{NL}=0$ is less than one percent.
\end{abstract}


\section{Introduction}
\label{sec:introduction}

Although the Cosmic Microwave Background (CMB) fluctuations are
usually assumed to follow a Gaussian distribution, the consensus is
that this is only an approximation valid to a certain level of
precision. The deviations from Gaussianity seem to be small and are
therefore difficult to estimate precisely. However, accurate knowledge
of both the size and nature of such a deviation would be of high value
in cosmology. The search for such anomalies in the CMB has therefore
attracted much attention recently.

Most models for inflation predict the CMB fluctuation to be slightly
non-Gaussian. This level of non-Gaussianity is measured by the
parameter $f_{NL}$ (see e.g.  \citep{inflationreview}). Alternative 
inflationary scenarios predict different values of $f_{NL}$, thus making
an accurate estimate of this parameter crucial for understanding the
physics of the inflationary era. Several $f_{NL}$ parameters have
been considered in the literature, corresponding to different ansatz
for the shape of primordial non-Gaussianities. In this paper we focus
on local non-Gaussianity, parametrized by $f_{NL}^{local}$ (simply
$f_{NL}$ in the following).

In \citep{rudjord}, we used needlets to estimate a value of $f_{NL}=
84 \pm 40$ (local type) different from zero at the $2\sigma$
level. This was in agreement with previous estimates
\citep{2008PhRvL.100r1301Y,2009ApJS..180..330K,smith2009,curto2008,curto2009,pietrobon2008,
  pietrobon2009} where values of $f_{NL}$ deviating from zero with
about $2\sigma$ were also reported. Also in agreement with previous
estimates, we found that the value of $f_{NL}$ increases with smaller
sky cuts. This could be a result of smaller error bars due to the
increased amount of data, or it could be an indication of foreground
residuals near the galactic plane influencing the value of
$f_{NL}$. This is our main motivation for estimating $f_{NL}$ on
smaller parts of the sky. By having local estimates of $f_{NL}$ we can
check whether the high value of $f_{NL}$ comes from areas close to the
galactic plane thus indicating the influence of foreground residuals,
or whether consistent $f_{NL}$ values are found in different parts of
the sky. The second motivation for studying the directional dependence
of $f_{NL}$, is due to several reports of the CMB deviating from
statistical isotropy in various ways
\citep{hansen2008,2009ApJ...690.1807G,hoftuft,eriksen2004,hansen2004,vielva2004,tegmark2003}. Here
we check whether a similar asymmetry is seen in the value of $f_{NL}$.

In \citep{rudjord}, the bispectrum of spherical needlets \citep{needbisp}
was used to obtain $f_{NL}$. The localization and uncorrelation properties of needlets
make them a convenient tool for studying localized regions on the
sky (see \citep{baldi2006}). We will here use a similar analysis to estimate the level of
non-Gaussianity on selected regions on the sky and in this way study
the spatial variations of $f_{NL}$. We therefore refer the reader to
(\cite{rudjord}) for a more detailed description of the procedure,
as well as for a more extensive list of references.

This paper is organized as follows. In section \ref{sec:data} we
describe the data used for the analysis, in section \ref{sec:maps} we describe the non-Gaussian maps and in section \ref{sec:method} we %briefly 
outline the method used for estimating $f_{NL}$. We apply
the procedure to \textit{WMAP} data and present the results in section
\ref{sec:results} before we summarize and conclude in section
\ref{sec:concl}.

 \section{Data}
\label{sec:data}
This analysis was performed using the foreground reduced co-added \textit{V+W}
frequency bands of the \textit{WMAP} five year data at
Healpix\footnote{http://healpix.jpl.nasa.gov} resolution $N_{side} =
512$. Also the individual Q, V and W bands were used for consistency tests.
The Gaussian simulations were generated using the best fit power
spectrum from the \textit{WMAP} five year release. We also used the
beam and noise properties supplied by the \textit{WMAP} team. For
masking out galactic foregrounds we used the \textit{KQ75} 
as well as the smaller \textit{KQ85} mask.

\section{Non-Gaussian maps}
\label{sec:maps}
The non-Gaussian maps used for the analysis have been generated using the algorithm developed in \citep{liguori:2003,liguori:2007} (see also \cite{ElsnerWandelt} for recent developments). 
We briefly review the general structure of the algorithm here, while addressing the reader to the above mentioned papers 
for further details. The CMB multipoles $a_{\ell m}$ are related to the primordial gravitational potential $\Phi$ 
through the well known formula:

\begin{equation}\label{eqn:phi2alm} 
a_{\ell m} = \int \frac{d^3 k}{(2 \pi)^3} \Phi(\mathbf{k}) Y_{\ell
  m}(\hat{k}) \Delta_\ell(k) \; ,
\end{equation}

where $\Delta_\ell(k)$ is the radiation transfer function and the potential is written in Fourier space. 
However, for the case of primordial non-Gaussianity, the primordial potential takes a very simple expression in 
real space, where:

\begin{equation}
\Phi(\mathbf{x}) = \Phi_L(\mathbf{x}) + f_{\rm NL} \left[\Phi_L^2(\mathbf{x}) - \left\langle \Phi_L^2(\mathbf{x}) \right \rangle \right] \; .
\end{equation}

In the previous expression $\Phi_{\rm L}$ is a Gaussian random field, characterized by a primordial power spectrum 
$P(k) = Ak^{n-4}$; in the following we will refer to $\Phi_{\rm L}(\mathbf{x})$ as the Gaussian part of the primordial 
potential. The remaining non-Gaussian part of the potential is simply the square of the Gaussian part in each point 
(modulo a constant term, necessary to enforce the condition $\langle \Phi(\mathbf{x}) \rangle = 0$; however it is clear 
that this term affects only the CMB monopole)

It is then convenient to work directly in real space and recast formula (\ref{eqn:phi2alm}) in the following form

\begin{equation}\label{eqn:phi2alm_real}
a_{\ell m} = \int d^3 x \Phi(\mathbf{x}) Y_{\ell m}(\hat{r}) \Delta_\ell(r) \; ,
\end{equation}
where $\Delta_{\ell}(r) \equiv \int dk \, k^2 j_\ell(kr) \Delta_{\ell}(k)$ are the real space transfer functions, $j_\ell(kr)$ 
is a spherical Bessel function, and 
$r$ is a lookback conformal distance. This
formula suggests to structure the algorithm in the following steps

\begin{enumerate}

\item Generate the Gaussian part $\Phi_L$ of the potential in a box whose side is the present cosmic horizon.

\item Square the Gaussian part point by point to get the non-Gaussian part. 

\item Expanding in spherical harmonics the Gaussian and non-Gaussian parts of the potential for different values 
of the radial coordinate $r$ in the simulation box.

\item Convolve the spherical harmonic expansions of $\Phi_{\rm L}$ and $\Phi_{\rm NL}$ with the radiation transfer function 
$\Delta_\ell(r)$ in order to obtain the Gaussian and non-Gaussian part of the multipoles of the final NG CMB simulation.
For a given choice of the non-Gaussian parameter $f_{\rm NL}$ a CMB map is then obtained simply through the linear 
combination $a_{\ell m} = a_{\ell m}^L + f_{\rm NL} a_{\ell m}^{\rm NL}$ (the superscripts L and NL always indicating Gaussian and 
non-Gaussian respectively).

\end{enumerate}

The most difficult and time consuming part in this process is actually the generation of the Gaussian part of the potential $\Phi$.
The difficulty arise from the fact that we are working in a box of the size of the present cosmic horizon, that in conformal time
is about $15$ Gpc, but at the same time a cell in this box must have a size no bigger than $20$ Mpc in order to resolve the last
scattering surface, where most of the CMB signal is generated. It turns out (see \cite{liguori:2003,liguori:2007} for details and 
more explanations) that a convenient way to achieve this is to work directly in spherical coordinates, use a non uniform 
discretization of the simulation box (since no points are needed in a large region of the box where photons are just free streaming) 
and generate the multipoles of the expansion of $\Phi_{\rm L}(\mathbf{x})$ through the following two step approach:

\begin{enumerate}

\item Generate uncorrelated radial multipoles $n_{\ell m}(r)$, gaussianly distributed and characterized by the following spectrum:

\begin{equation}
\label{eqn:whitenoise} 
\left \langle n_{\ell_1 m_1}(r_1)
n^*_{\ell_2 m_2}(r_2) \right \rangle = 
\frac{\delta^D(r_1-r_2)}{r^2}\delta_{\ell_1}^{\ell_2} \delta_{m_1}^{m_2}\; ;  
\end{equation} 

where $\delta^D$ is the Dirac delta function. 
\item Filter the multipoles $n_{\ell m}$ with suitable functions in order to produce a Gaussian random field with the properties 
of the multipole expansion of the primordial Gaussian potential $\Phi_L$ . It can be shown that the expression of the filter functions
is:

\begin{equation}
\label{eqn:filter} 
W_\ell(r,r_1) =
\frac{2}{\pi} \int \! dk \, k^2 \, \sqrt{P_\Phi(k)} \, j_\ell(kr)
j_\ell(kr_1) \; ,  
\end{equation}

where $P_{\Phi}$ is the primordial power spectrum, and the filtering operation takes the form 

\begin{equation}
\label{eqn:nlm2phil} 
\Phi^{\rm L}_{\ell m}(r) = \int \! dr_1 \, r_1^2 \, n_{\ell m}(r_1) 
W_\ell(r,r_1) \; .
\end{equation} 

In the last expression $\Phi^{\rm L}_{\ell m}(r)$ are the desired quantities i.e. the multipoles of the expansion of the Gaussian 
part of the primordial potential for a given $r$.

\end{enumerate}

\section{Method}
\label{sec:method}
The bispectrum has shown to be the most powerful tool for estimating
$f_{NL}$. The bispectrum is zero for a Gaussian field, and any
significant deviations from a zero bispectrum is therefore a
non-Gaussian signal. 
Needlets are a new type of spherical wavelets which were introduced by
\citep{npw1}. For our analysis we use the bispectrum of needlet
coefficients \citep{needbisp} to estimate $f_{NL}$. The localization
properties of needlets make it possible to obtain the bispectrum in
several different regions of the sky with very little additional costs
in terms of CPU time compared to one full sky analysis. Although the
needlet bispectrum does not yield optimal error bars on $f_{NL}$ (see \citep{smith2009}
for the optimal method), the
advantage is the possibility of a fast and easy calculation of local
estimates for consistency checks.

The needlet coefficients are denoted $\beta_{jk}^B$, $j$ is frequency,
$k$ the direction on the sky (we will take $k$ as the pixel number in
the Healpix grid) and the parameter $B$ characterizes the localization
in frequency domain.  Indeed, needlets allow for a tight control of
localization in harmonic space and uncorrelation in pixel space; these
properties are valuable for statistical inference and are not shared
by other wavelet constructions, please refer to \citep{baldi2006}),
\citep{mpbb08} for details and further references (see also
\citep{pietrobon2006} and \citep{fay08}).  For instance, for a given
value of $B$ a needlet coefficient only contains information on
multipoles in the range $\ell=[B^{j-1},B^{j+1}]$. Thus, the parameter
$B$ controls localization in harmonic space: small values of $B$
correspond to small ranges of frequencies $j$, while the reverse is
true for larger $B$.

The needlet bispectrum with base $B$ may be expressed as
\begin{equation}
\label{eq:needbisp}
I_{j_1 j_2 j_3}(\mathrm d \Omega) = \sum_k^{\mathrm{pixels} \in  \mathrm d \Omega} \frac{\beta_{j_1 k} \beta_{j_2 k} \beta_{j_3 k}}{\sigma_{j_1 k} \sigma_{j_2 k} \sigma_{j_3 k}}
\end{equation}
where $\sigma_{j k}^B$ are the standard deviation of the needlet
coefficients $\beta_{jk}^B$ for a Gaussian map and $\mathrm d \Omega$ is the region of the sky for which the bispectrum is calculated.

For the estimation of $f_{NL}$ we used a similar procedure as in \cite{rudjord}, but with a few small changes. We used a higher value for $B$, resulting in 
fewer needlet scales. Intuitively one would expect this to give a poor result, since 
each needlet scale would cover a large interval in $\ell$ -space. However, 
a higher value for $B$ also gives better localization properties in pixel-space, thus 
minimizing the influence of the mask and thereby reducing the error bars. Additionally, this greatly reduces the computational 
cost of the analysis.

 We will calculate the needlet bispectra using the
needlet coefficients calculated on the pixels $k$ of a
$N_\mathrm{side}=512$ map. We will use all the $N_\mathrm{side}=512$
pixels $k$ which are inside an $N_\mathrm{side}=2$ pixel as our
smallest region $\mathrm d \Omega$. We thus obtain 48 bispectra
$I_{j_1 j_2 j_3}(p)$ where $p$ denotes a pixel of the
$N_\mathrm{side}=2$ pixelization. Having obtained these 48 bispectra
it is straightforward to construct the bispectrum on sky patches with
different shapes and sizes. We see from eq. \ref{eq:needbisp}
that the bispectrum of a larger patch is then simply
\begin{equation}
\label{eq:needbispadd}
  I_{j_1 j_2 j_3}(\mathrm{region}) = \sum_pI_{j_1 j_2 j_3}(p),
\end{equation}
where the sum over $p$ goes over the $N_\mathrm{side}=2$ pixels within
the desired region.

In order to estimate $f_{NL}$ we perform a $\chi^2$ analysis for every
map and region to be investigated
\begin{equation}
  \chi^2(f_{NL}) = \mathbf d^T(f_{NL}) \mathbf C^{-1} \mathbf d(f_{NL})
\end{equation}
where the data vector is
\begin{eqnarray}
  \mathbf d &=& I^{\mathrm{obs}}_{j_1 j_2 j_3} - \langle I_{j_1 j_2 j_3}(f_{NL})\rangle \notag \\
  &=& I^{\mathrm{obs}}_{j_1 j_2 j_3} - f_{NL}\langle \hat I_{j_1 j_2 j_3} \rangle .
\end{eqnarray}
$\langle \hat I_{j_1 j_2 j_3} \rangle$ is here the average first order
non-Gaussian bispectrum obtained using non-Gaussian simulations \citep{liguori:2007}. See \citep{rudjord} for details. The
corresponding covariance matrix $\mathbf C$ is evaluated by means of
Monte-Carlo simulations.

Differentiating to find the value for $f_{NL}$ which gives the lowest
$\chi^2$ yields the ``Generalized Least Squares'' estimate:
\begin{equation}
  f_{NL} = \frac{\left<\hat I_{j_1 j_2 j_3}\right>^T \mathbf{C}^{-1} I_{j_1 j_2 j_3}}{\left<\hat I_{j_1 j_2 j_3}\right>^T \mathbf{C}^{-1}\left<\hat I_{j_1 j_2 j_3}\right>} \label{eq:the_fnl_equation}.
\end{equation}

We estimate local $f_{NL}$ values according to the procedure described in \citep{rudjord}, but with one important difference: We calculate and save the bispectra calculated on each individual $N_{side}=2$ pixel for each simulation. Then, when the bispectrum and thereby $f_{NL}$ is estimated for a larger region, the bispectra for the different $N_{side}=2$ pixels are added up according to eq. \ref{eq:needbispadd} and the correlation matrix is constructed from this final bispectrum for each larger region.

\section{Results}
\label{sec:results}
First we estimated $f_{NL}$ on the full sky using $B=1.781$ and $11$ needlet scales
covering multipoles up to $\ell_{max} = 1500$. This was the best trade-off (lowest error bars) between needlet coefficients for low values of $B$ being more affected by the mask but having more frequencies $j$ and for high values of $B$ being less affected by the mask but having fewer scales. Otherwise we followed the same procedure as in
\cite{rudjord}. The results are presented in table \ref{table:fullsky}.

\begin{table}[htdp]
\begin{center}
\begin{tabular}{|l|c|c|}
  \hline
  freq. channel  & \textit{KQ75} & \textit{KQ85} \\
  \hline
  \textit{V} + \textit{W} & $73 \pm 31$  & $78 \pm 29$ \\
  \textit{V}     & $58 \pm 35$ & $55 \pm 33$ \\
  \textit{W}     & $74 \pm 37$ & $72 \pm 34$ \\
  \textit{Q}     & $-8 \pm 39$ & $-9 \pm 37$  \\
  \hline
\end{tabular}
\end{center}
\caption{$f_{NL}$ estimates and $1 \sigma$ error bars. These estimates are found using the full CMB sky.}
\label{table:fullsky}
\end{table}

As we see from the results, the combined \textit{V} + \textit{W}
channel gives a $2 \sigma$ deviation from Gaussianity. The individual
\textit{V} and \textit{W} channels are consistent (within $2\sigma$), but the \textit{Q}
channel deviates significantly from the others, suggesting possible
contamination by foregrounds. 

We repeat the above analysis on localized regions on the sphere. First
the bispectra were found for an ensemble of Gaussian simulations on
each of the $48$ $(N_{side}=2)$ Healpix patches on the sky. We then
combined these patches in three different ways according to
eq. \ref{eq:needbispadd}.
\begin{enumerate}
\item \textbf{Hemispheres} The larger regions were defined to be
  hemispheres. For each of the 48 directions defined by the
  $(N_{side}=2)$ pixel centers, $f_{NL}$ was estimated on a hemisphere
  centered on this direction.
\item \textbf{$45^\circ$ discs} Same as for the hemispheres,
  estimating $f_{NL}$ on discs with $45^\circ$ radius instead of
  hemispheres.
\item \textbf{Equatorial rings} These regions were defined
  to be the $7$ constant latitude rings in which the
  $N_\mathrm{side}=2$ healpix pixels are ordered. Each of these rings
  covers a relatively small part of the sky, and the error-bars are
  therefore large. However they are not overlapping, and the $f_{NL}$
  estimates on the different rings can therefore be considered
  nearly independent (except for the largest scales).
\end{enumerate}

\begin{figure}
  \begin{center}
    \includegraphics[width=8cm]{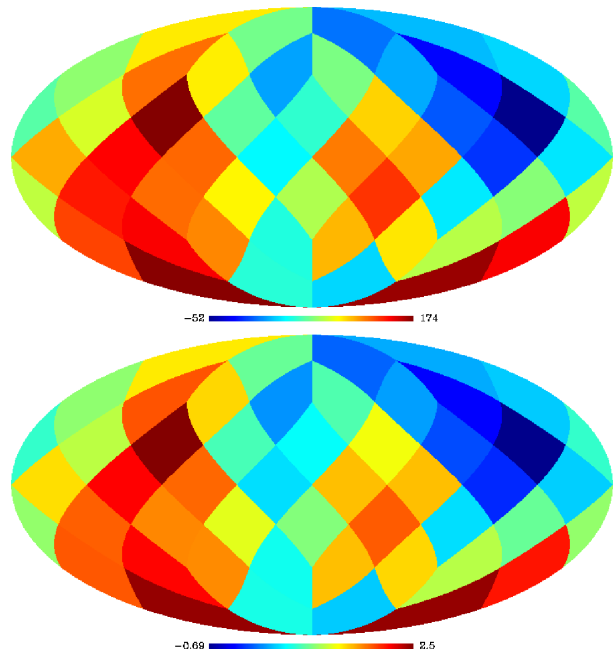}
  \end{center}
  \caption{The upper figure shows $f_{NL}$ estimates on $45^\circ$ discs
    centered on the given pixels, while the lower figure shows the
    same estimates divided by their standard deviation. The estimates were made on the V+W channel using the KQ85 galactic mask.}
 \label{fig:fnl_sky}
\end{figure}

\begin{figure}
  \begin{center}
    \includegraphics[width=8cm]{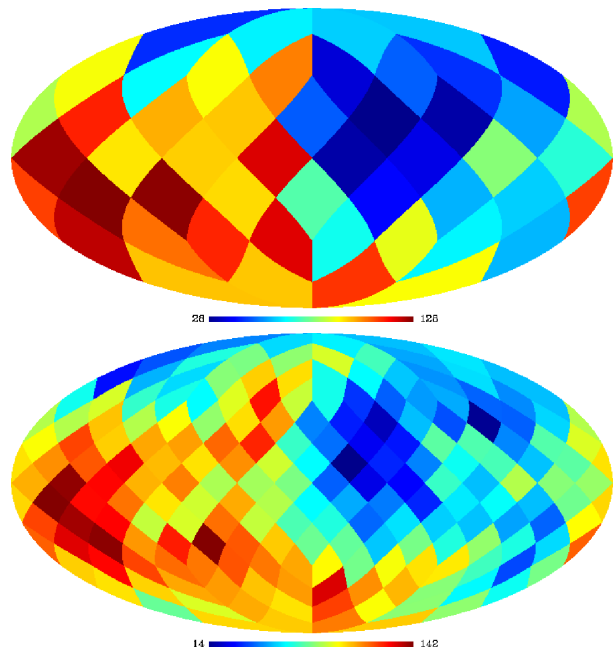}
  \end{center}
  \caption{Both the figures shows $f_{NL}$ estimates on hemispheres,
    but for the upper one the estimates are performed while using
    resolution $N_{side} = 2$, while for the lower one we have used
    $N_{side} = 4$. We can see that the two maps show the same
    structures. These estimates were made on the V+W channel using
    the KQ75 galactic mask.}
 \label{fig:fnl_sky2}
\end{figure}

The $f_{NL}$ estimates from the $45^\circ$ discs with the smaller
\textit{KQ85} mask are shown in figure \ref{fig:fnl_sky}, where each
pixel represents the $f_{NL}$ estimate on a disc centered on the
pixel. The lower plot shows the same map, but $f_{NL}$ for each region
has been normalized by its standard deviation. We see that even for
the smaller mask, there is no evidence for particularly high values of
$f_{NL}$ in the galactic region, but note that most of the values are
positive. In figure \ref{fig:fnl_sky2} we show the corresponding map
of $f_{NL}$ estimates over hemispheres while using the \textit{KQ75}
mask. As an additional test of consistency we have produced the same estimates based on bispectra obtained from $N_\mathrm{side}=4$ pixels instead of $N_\mathrm{side}=2$ pixels. The result is shown in the same figure. We can see that the two maps show the same structures on the sky.

As seen from figure \ref{fig:fnl_sky} the distribution of $f_{NL}$ on
the sky shows a dipole. For the
$f_{NL}/\sigma$ map of the $45^\circ$ discs of the \textit{V+W} channel (KQ85),
the dipole has a maximum at $\theta=129^\circ$, $\phi=96^\circ$ with an amplitude of $T_{\ell = 1}^{f_{NL}} =
61$. We investigated whether such a dipolar distribution of $f_{NL}$ was common in Gaussian 
simulations and found the value of the dipole amplitude to be in good agreement with simulated maps.
The dipole is therefore to be expected.  Investigation of the
hemisphere results as well as for the \textit{KQ75} mask using both the combined \textit{V+W}
channel and the individual \textit{Q},
\textit{V} and \textit{W} frequency channels yielded similar results.

In order to check whether the 48 estimates of $f_{NL}$ were internally
consistent, we combined the estimates to form a data vector $\mathbf d
=[f_{NL}^1,f_{NL}^2,...,f_{NL}^{48}]$ for a $\chi^2$ test. Simulations
were then used to find a covariance matrix containing the covariances
between the 48 estimates. Then a $\chi^2$ test was performed on
Gaussian simulations as well as on the \textit{WMAP} data. The
$\chi^2$ values of the data were compared with those of the
simulations. The results, both for the hemisphere estimates as well as
for the $45^\circ$ disk estimates, was that the $\chi^2$ of the
\textit{WMAP} data was fully consistent with Gaussian simulations (within $2\sigma$). The local $f_{NL}$ estimates are therefore internally consistent.

As a further test we obtained estimates of $f_{NL}$ on equatorial
rings. The motivation behind this approach was to uncover possible
foreground contamination outside the \textit{KQ75} mask. The $f_{NL}$
estimates for the different rings are presented in table
\ref{table:results_rings}. We see that the estimates of $f_{NL}$ seem
higher around equator, but the errors-bars are also larger (due to the
galactic mask in the equatorial region).  Since none of the rings show
particularly high values (compared to the error bars) we do not have
evidence to claim that foreground residuals have an influence of the
estimates of $f_{NL}$. The internal consistency of the various ring
estimates have been tested as described above and shown to be in
agreement with simulations (within $2\sigma$).

We also estimated $f_{NL}$ on the equatorial rings of the \textit{V +
  W} map using the smaller \textit{KQ85} mask. These results are
presented in table \ref{table:results_rings}.  As expected the
error-bars are somewhat smaller then the results with the
\textit{KQ75} mask, especially around the equatorial region. The
estimates are consistent with the ones from the \textit{KQ75} mask.

We then followed through with similar investigations of the individual
\textit{Q}, \textit{V} and \textit{W} frequency channels using the
\textit{KQ75} mask (table \ref{table:results_rings}). The \textit{V}
and \textit{W} channels are consistent (within $2\sigma$), while only the \textit{Q}
channel shows a $3\sigma$ deviation in the ring around equator exactly
where foreground residuals would be expected. We thus suspect a
possible influence of foreground contamination in this band. This was
further checked by testing the consistency of the estimated $f_{NL}$
between the bands using a $\chi^2$ approach. Whereas the (V-W) and
(V-Q) differences were found to be consistent with simulations (within $2\sigma$), the
(W-Q) difference was found to be larger than in $99\%$ of the
simulations.

Also, a similar test was performed on the differences
$f_{NL}(V)-f_{NL}(VW)$ and $f_{NL}(W)-f_{NL}(VW)$. We see in table
\ref{table:fullsky} and on some rings in table
\ref{table:results_rings} that the VW estimates seem driven by the W
estimate. This is not seen in simulated maps and it was found that the
small difference $f_{NL}(W)-f_{NL}(VW)$ found for the WMAP data is
found in only $5\%$ of the simualted maps whereas the difference
$f_{NL}(V)-f_{NL}(VW)$ for WMAP was found to be consistent (well
within $2\sigma$) with simulations.

\begin{table}[htdp]
\begin{center}
\begin{tabular}{|l|r|r|r|r|r|}
  \hline
  & \textit{V+W} & \textit{V+W} & \textit{Q} & \textit{V} & \textit{W} \\
  \hline
  ring& \textit{KQ75} & \textit{KQ85} & \textit{KQ75} & \textit{KQ75} & \textit{KQ75} \\

  \hline
  $1$ & $91\pm 95$ & $93 \pm 95$ & $47 \pm 118$ & $66 \pm 106$  & $94 \pm 111$ \\
  $2$  & $11\pm 68$ & $6 \pm 68$& $-18 \pm 83$  & $1 \pm 76$    & $28 \pm 79$  \\
  $3$ & $80\pm 80$& $43 \pm 71$ & $-149 \pm 100$ & $19 \pm 90$   & $13 \pm 93$  \\
  $4$ & $283\pm 183$ & $122 \pm 113$& $700 \pm 226$ & $462 \pm 205$ & $253 \pm 213$  \\
  $5$ & $117\pm 82$ & $128 \pm 70$& $-61 \pm 103$  & $37 \pm 92$   & $122 \pm 96$  \\
  $6$  & $39\pm 66$& $53 \pm 66$ & $-81 \pm 81$  & $35 \pm 74$   & $15 \pm 78$ \\
  $7$  & $158\pm 93$ & $156 \pm 93$& $174 \pm 114$ & $138 \pm 104$ & $201 \pm 108$  \\
  \hline
\end{tabular}
\end{center}
\caption{The $f_{NL}$ estimates and $1 \sigma$ error-bars for equatorial rings 
  of the individual \textit{Q}, \textit{V}, and \textit{W} frequency channels as well 
  as the co-added \textit{V+W} for the masks \textit{KQ75} and \textit{KQ85}.}
\label{table:results_rings}
\end{table}

We see that all of the rings for the \textit{V+W} channel (as well as
the individual \textit{V} and \textit{W} channels) give a estimate
$f_{NL} \ge 0$. These rings are nearly independent (except for the largest
scales) and for a sky with $f_{NL} = 0$ one would expect that each of
these have a $50\%$ probability of being positive. For $7$ rings one
would then expect only a $\frac{1}{2^{7}} \approx 0.78\%$ probability
that all rings give $f_{NL} \ge 0$ (this probability is confirmed by
simulations).

Since the lowest value estimated for the rings of the \textit{V + W}
channel (with the \textit{KQ75} mask) is $f_{NL} = 11$, we
investigated the probability of this occurring in Gaussian
simulations.  We found that of $4000$ Gaussian simulations, only
$0.35\%$ have $f_{NL} \ge 11$ in all the rings.

\section{Conclusions}
\label{sec:concl}
In this paper we have used the bispectrum of needlets to obtain local
estimates of $f_{NL}$ on the WMAP five year data. We performed the
analysis on the combined \textit{V+W} channel, as well as the
individual \textit{Q}, \textit{V} and \textit{W} channels, using
multipoles up to $\ell = 1500$ and the \textit{KQ75} galactic cut. For
the combined \textit{V+W} channel we also applied the \textit{KQ85}
mask.

We first made a full sky analysis, resulting in a best fit value of
$f_{NL} = 73 \pm 31$ for the combined \textit{V+W} channel using the
\textit{KQ75} mask. The individual \textit{V} and \textit{W} channels
give consistent (within $2\sigma$) results, but the $f_{NL}$ estimate of the \textit{Q}
channel deviates significantly, suggesting contamination of
foregrounds.

The estimates were then made on selected regions of the sky and showed
how the needlet bispectrum approach is powerful for finding estimates
of $f_{NL}$ in many different regions, roughly at the cost of one
single full sky estimate. We divided the sky into smaller regions
according to four different patterns: hemispheres, $45^\circ$ disks
and equatorial rings. In each of these schemes $f_{NL}$ was estimated
in every region. The results were compared to simulations using a
$\chi^2$ test, and all local $f_{NL}$ estimates were found to be
internally consistent (within $2\sigma$) for the \textit{V} and \textit{W} channels.

The local estimates of $f_{NL}$ showed a dipolar distribution
of $f_{NL}$ on the sphere, with a maximum at $\theta = 129^\circ$,
$\phi = 96^\circ$. For comparison, the hemispherical power asymmetry
reported in \cite{hansen2008} was found with a maximum in $\theta =
107^\circ$, $\phi = 226^\circ$, and is therefore not expected to be
connected to the findings in this paper.

Also, such a dipolar distribution was found to be
common in simulated Gaussian maps. For the equatorial rings we found a positive
value for $f_{NL}$ in every ring. We compared the lower estimate of
the \textit{V+W} channel ($f_{NL} \ge 11$ for all rings) with
simulations and found that only $0.35 \%$ of Gaussian simulations have
this feature.

For the rings we find no significant evidence of
foreground contamination outside the galactic \textit{KQ75} mask. The
results also seem to be fairly consistent (within $2\sigma$) between the individual
\textit{V} and \textit{W} channels whereas the \textit{Q} band show 
signs of possible foreground contamination in the equatorial band 
where $f_{NL}$ is larger than zero at the $3\sigma$ level. This is 
confirmed by the fact that the differences in local $f_{NL}$ values 
between the  \textit{Q} and  \textit{W} bands are larger than in 
$99\%$ of the simulations.

We conclude that our study shows no significant anisotropy in the
estimates of $f_{NL}$ in the CMB sky. No abnormal values for $f_{NL}$ 
are found close to the equator except for the  \textit{Q} band where 
we suspect foregrounds to influence the estimate of  $f_{NL}$.

\begin{acknowledgements}
  FKH is grateful for an OYI grant from the Research Council of
  Norway.  This research has been partially supported by ASI contract
  I/016/07/0 "COFIS" and ASI contract Planck LFI Activity of Phase
  E2. We acknowledge the use of the NOTUR supercomputing
  facilities. We acknowledge the use of the HEALPix
  \citep{gorski:2005} package and the Legacy Archive for Microwave
  Background Data Analysis (LAMBDA). Support for LAMBDA is provided by
  the NASA Office of Space Science.
\end{acknowledgements}

\end{document}